\documentclass[journal]{IEEEtran}

\ifCLASSINFOpdf
\else
\fi

\usepackage{setspace}
\usepackage{bbm}
\usepackage[cmex10]{amsmath}
\usepackage{amssymb}
\usepackage{cite}
\usepackage{graphicx}
\usepackage{array,color}
\usepackage{amsmath}
\allowdisplaybreaks
\usepackage{stfloats}
\usepackage{graphicx}
\usepackage{epstopdf}
\usepackage{subfigure}
\usepackage{tabularx}
\usepackage{epsfig,epsf,color,balance,cite}
\usepackage{algorithmic}
\usepackage{algorithm}
\usepackage{url}
\usepackage{bm}

\usepackage{amsthm}
\ifodd 0
\usepackage{soul}
\usepackage{color}
\setstcolor{red}

\newcommand{\rev}[1]{{\color{red}#1}} 
\newcommand{\del}[1]{\st{#1}} 

\newcommand{\com}[1]{\textbf{\color{red} (COMMENT: #1)}} 
\newcommand{\response}[1]{\textbf{\color{green} (RESPONSE: #1)}} 
\else

\newcommand{\rev}[1]{#1}

\newcommand{\del}[1]{}

\newcommand{\com}[1]{}
\newcommand{\comg}[1]{}
\newcommand{\response}[1]{}
\fi

\title{\huge {Intelligent Reflecting Surface-Assisted Multiple Access with User Pairing: NOMA or OMA?} }

\author{\vspace{-0.2cm} Beixiong Zheng,~\IEEEmembership{Member,~IEEE}, Qingqing Wu,~\IEEEmembership{Member,~IEEE}, and Rui Zhang,~\IEEEmembership{Fellow,~IEEE} \vspace{-0.85cm}
	\thanks{\vspace{-0.5cm}
		
		The authors are with the Department of Electrical and Computer Engineering, National University of Singapore,
		email: \{elezbe, elewqq, elezhang\}@nus.edu.sg.

	}
}

\begin{document}
\markboth{IEEE Communications Letters, Vol. XX, No. XX, XXX 2020}{SKM: My IEEE article}
\maketitle

\begin{abstract}
	The integration of intelligent reflecting surface (IRS) to multiple access networks is 
	a cost-effective solution for boosting spectrum/energy efficiency and enlarging network coverage/connections. However, due to the new capability of IRS in reconfiguring the wireless propagation channels,
	it is fundamentally unknown which multiple access scheme is superior in the IRS-assisted wireless network.
	In this letter, we pursue a theoretical performance comparison between non-orthogonal multiple access (NOMA) and orthogonal multiple access
	(OMA) in the IRS-assisted downlink communication, for which the transmit power minimization problems are formulated under the \emph{discrete unit-modulus} reflection constraint on each IRS element.
	\rev{We analyze the minimum transmit powers required by different multiple access schemes and compare them numerically, which turn out to not fully comply with the \emph{stereotyped superiority} of NOMA over OMA in conventional systems without IRS.}
	Moreover, to avoid the exponential complexity of the brute-force search for the optimal discrete IRS phase shifts, we propose a low-complexity solution to achieve near-optimal performance.


\end{abstract}
\begin{IEEEkeywords}
	Intelligent reflecting surface (IRS), non-orthogonal multiple access (NOMA),
	frequency division multiple access (FDMA), time division multiple access (TDMA),
	user pairing,
	power minimization, discrete phase shifts.
\end{IEEEkeywords}
\IEEEpeerreviewmaketitle
\vspace{-0.4cm}
\section{Introduction}
\IEEEPARstart{I}{ntelligent}
 reflecting surface (IRS) has recently attracted growing attention and is envisioned as an innovative technology for the beyond fifth-generation (B5G) communication system, due to its potential of achieving significant improvement in communication coverage, throughput, and energy efficiency \cite{qingqing2019towards,Renzo2019Smart,Wu2019TWC}.
Specifically, IRS is a planar meta-surface composed of a large number of
reconfigurable passive elements, which are attached with
a smart controller to enable dynamic adjustment on the signal reflections for different
purposes, such as signal power enhancement and interference suppression.
In particular, compared to conventional techniques such as active relaying/beamforming, 
IRS not only reflects signals in a full-duplex and noise-free manner without incurring self-interference,
but also greatly saves energy consumption and hardware/deployment cost by using lightweight passive components only \cite{qingqing2019towards,Wu2019TWC}.


On the other hand, non-orthogonal multiple access
(NOMA) has also received significant attention
and shown superiority over orthogonal multiple access (OMA) in conventional wireless systems without IRS,
 for improving the spectral efficiency, balancing user fairness, and enlarging network connections.
In the downlink NOMA, the user of stronger channel with the base station (BS) or access point (AP) employs the successive interference cancellation
(SIC) technique to cancel the co-channel interference from the users of weaker channels, prior to decoding its own message. \rev{As a result, the decoding order depends on user channel power gains, which are determined by the propagation environments and user locations.
In contrast, since IRS is capable of reconfiguring user channels by controlling the reflected signal amplitudes and/or phase shifts,
the user decoding order of NOMA can be \emph{permuted} by adjusting the IRS reflection to achieve more flexible performance trade-offs among the users.}
Nevertheless, for IRS to be integrated into the future wireless network, it still remains unknown whether NOMA is always superior to OMA as in conventional systems without IRS.
Although some relevant works on the application of IRS to NOMA have recently appeared \cite{Gang2019intelligent,ding2019simple,li2019joint,mu2019exploiting},
the theoretical performance comparison between NOMA and OMA for IRS-assisted wireless communications is not well understood yet, to the best of our knowledge.



\begin{figure}[!t]
	\centering
	\includegraphics[width=2.3in]{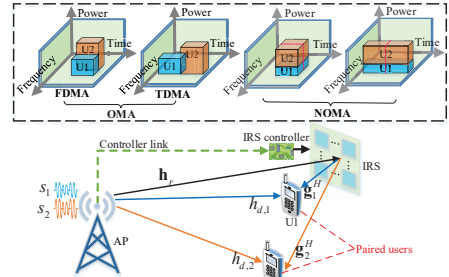}
	\caption{An illustration of OMA and NOMA in the IRS-assisted system.}
	\vspace{-0.6cm}
	\label{system}
\end{figure}
To address the above issue, in this letter we investigate the theoretical performance comparison between NOMA and OMA in an IRS-assisted downlink communication system, where 
two types of OMA schemes, i.e., frequency division multiple access (FDMA) and time division multiple access (TDMA), are considered, as shown in Fig.~\ref{system}.
The problems for minimizing the transmit power
at the AP are formulated for both OMA and NOMA, subject to the \emph{discrete unit-modulus reflection} constraint at the IRS and the users' target rates.
By analyzing the relationship of minimum transmit powers required by different multiple access schemes, we unveil that the minimum transmit power of FDMA is always no lower than that of the other two counterparts (TDMA and NOMA), \rev{while the minimum power comparison between NOMA and TDMA depends on the target rates and locations of the users. Specifically, TDMA requires lower transmit power than NOMA when two near-IRS users are paired and have symmetric rates, whereas the relationship is reversed otherwise.}
Moreover, to avoid the exponentially high complexity of the brute-force search for the optimal discrete IRS phase shifts, we propose an efficient low-complexity algorithm based on the linear approximation initialization to effectively balance the user channel power gains for minimizing the transmit power, followed by the alternating optimization to achieve near-optimal performance.
\vspace{-0.3cm} 
\section{System Model and Problem Formulation}

As depicted in Fig.~\ref{system}, we consider an IRS-assisted downlink communication system, 
where an IRS is deployed to assist in the transmission
from a single-antenna AP to multiple single-antenna users.
\rev{Similar to \cite{zheng2019intelligent,yang2019intelligent}, an IRS composed of $N$ reflecting elements
is divided into $M$ sub-surfaces, each of which consists
of ${\bar N}= N/M$ adjacent elements to share a common reflection coefficient for reducing the implementation complexity.
Moreover, the passive IRS is connected to a smart controller that enables dynamic adjustment on the reflections of IRS elements and also plays the role of exchanging information between the IRS and AP via a separate wireless link \cite{qingqing2019towards,Wu2019TWC}.}
For ease of exposition, \rev{we focus on the case with any two users sharing two given adjacent time-frequency resource blocks (RBs) in practice, where the channels are assumed to be frequency-flat fading and constant over the two adjacent time-frequency RBs (see Fig.~\ref{system}).
It is worth pointing out that by adopting the concept of user pairing, the two-user case can be extended to the general multi-user case.}
Without loss of generality, it is assumed that user~1 is located in the vicinity of the IRS while user~2 can be arbitrarily located within the coverage of the AP (i.e., in or out of the IRS's coverage).
As shown in Fig.~\ref{system}, we consider two types of OMA schemes, i.e., FDMA and TDMA, which serve two users over two adjacent RBs separated in the frequency and time domains, respectively;
while the NOMA scheme serves two users simultaneously over the two adjacent RBs either in the frequency- or time-domain, for fair comparison with OMA.

To characterize the optimal performance of different multiple access schemes in the IRS-assisted system, we assume perfect channel state information (CSI) available at the AP. Note that the channel estimation methods proposed in \cite{qingqing2019towards,zheng2019intelligent} can be applied to acquire the CSI of the two users independently.
The baseband equivalent channels from the AP to IRS, from the IRS to user $k$, and from the
AP to user $k$ are denoted by ${\bm h}_r \in {\mathbb{C}^{M\times1 }} $, ${\bm g}^H_k\in {\mathbb{C}^{1\times M}}$, and $h_{d,k}\in {\mathbb{C}}$, respectively, with $k\in \{1,2\}$.
Let ${\bm \theta}\triangleq[{\theta_1},  {\theta_2},\ldots,{\theta_M}]^T=[\beta_1 e^{j \phi_1}, \beta_2 e^{j \phi_2},\ldots,\beta_M e^{j \phi_M}]^T$ denote the equivalent reflection coefficients of the IRS, where $\phi_m \in [0, 2\pi)$ and $\beta_m \in [0, 1]$ are
the phase shift and reflection amplitude of the $m$-th sub-surface, respectively.
To maximize the signal power reflected by the IRS and reduce the hardware cost,
we set $\beta_m=1, \forall  m=1,\ldots,M$ and \rev{consider the practical discrete unit-modulus constraint ${\cal F}=\left\{e^{j \phi}\big|\phi\in \{0, \Delta\phi,\ldots, (L-1)\Delta\phi\}\right\}$ for each phase shift coefficient $\theta_m$,} where $\Delta\phi=2\pi/L$ with $L$ denoting the number of discrete phase-shift levels \cite{wu2019beamforming}.

\vspace{-0.25cm}
\subsection{NOMA Transmission Scheme}
As shown in Fig.~\ref{system}, the AP \emph{simultaneously} transmits the signals of two users by adopting the superposition coding over two frequency/time domain adjacent RBs, which is given by
\vspace{-0.15cm}
\begin{align}
x= \sqrt{P_1 } {s_{1}} + \sqrt{P_2}{s_{2}}
\end{align}
where $P_k$ denotes the power allocated
to user $k$ with $k\in \{1,2\}$ and $s_k$ denotes the transmitted data symbol for user~$k$, which is assumed to be a circularly symmetric complex Gaussian (CSCG) random variable with zero mean and unit variance.
Accordingly, the signal received at user $k$ is given by
\vspace{-0.2cm}
\begin{align}\label{useR_signal}
\hspace{-0.2cm}y_k\hspace{-0.1cm}=\hspace{-0.1cm} \left(\hspace{-0.05cm}  {\bm g}^H_k {\bm \Theta} {\bm h}_r  \hspace{-0.05cm}+\hspace{-0.05cm} h_{d,k}\hspace{-0.05cm}\right)  \left(\hspace{-0.1cm}\sqrt{P_1 } {s_{1}} \hspace{-0.05cm}+\hspace{-0.05cm} \sqrt{P_2}{s_{2}}\hspace{-0.1cm}\right)+n_k, k\in \{1,2\}
\end{align}
where $n_k$ denotes the additive CSCG noise with zero mean and variance $\sigma^2$ at user $k$ and ${\bm \Theta}=\text{diag} \left( {\bm \theta} \right)$ represents \rev{the diagonal phase-shift matrix of the IRS.
By denoting ${\bm q}^H_k\triangleq {\bm g}^H_k \text{diag} \left({\bm h}_r \right)$ as the cascaded AP-IRS-user channel before the IRS phase-shift adjustment, \eqref{useR_signal} can be rewritten as}
\vspace{-0.15cm}
\begin{align}\label{useR_signal2}
\hspace{-0.2cm}y_k \hspace{-0.05cm}=\hspace{-0.05cm}  \left(  {\bm q}^H_k {\bm \theta}   + h_{d,k}\right)  \left(\sqrt{P_1 } {s_{1}} \hspace{-0.05cm}+\hspace{-0.05cm} \sqrt{P_2}{s_{2}}\right)+n_k, k\in \{1,2\}.
\end{align}
Let $\lambda_1({\bm \theta})=|{\bm q}^H_1 {\bm \theta}+h_{d,1}|^2$ and $\lambda_2({\bm \theta})=|{\bm q}^H_2 {\bm \theta}+h_{d,2}|^2$ denote the channel power gains of users 1 and 2, respectively. \rev{Note that since channel power gains $\lambda_1({\bm \theta})$ 
and $\lambda_2({\bm \theta})$ are two discrete functions which vary with the IRS phase-shift vector ${\bm \theta}$, we may have $2!=2$ permutations for the user decoding order.}
Given the target rates of the two users, we aim to minimize the total transmit power
at the AP by optimizing \rev{the IRS phase-shift vector ${\bm \theta}$,} subject to the discrete unit-modulus constraints of IRS elements. The
corresponding optimization problem can be decomposed into two sub-problems associated with two different decoding orders as follows.
\vspace{-0.15cm}
\begin{align}
\text{(N1):}~
P_{N1}\triangleq& \underset{{\bm \theta},P_1, P_2}{\text{min}}
& &  P_1+P_2\\
& \text{~~s.t.} & & \log_2\left(1+\frac{P_1 \lambda_1({\bm \theta}) }{\sigma^2} \right)\ge \gamma_1\label{con0_N1}\\
& & & \log_2\left(1+\frac{P_2 \lambda_2({\bm \theta}) }{\sigma^2+P_1  \lambda_2({\bm \theta}) } \right) \ge \gamma_2\label{con1_N1}\\
& & & \theta_m \in {\cal F}, \forall m=1,\ldots,M\\
\text{(N2):}~
P_{N2}\triangleq& \underset{{\bm \theta},P_1, P_2}{\text{min}}
& & P_1+P_2\\
& \text{~~s.t.} & & \log_2\left(1+\frac{P_1 \lambda_1({\bm \theta}) }{\sigma^2+P_2  \lambda_1({\bm \theta}) } \right)\ge \gamma_1  \label{obj_N2}\\
& & & \log_2\left(1+\frac{P_2 \lambda_2({\bm \theta}) }{\sigma^2} \right)\ge \gamma_2\label{con1_N2}\\
& & & \theta_m \in {\cal F}, \forall m=1,\ldots,M
\end{align}
where $\gamma_1$ and $\gamma_2$ are the target rates of users 1 and 2 in bits per second per Hertz (bps/Hz), respectively. 
Since the user rates are monotonically increasing with $P_1$ and $P_2$, 
the inequality rate constraints should be met with equality at the optimal solution.
By eliminating the equality constraints, (N1) and (N2) can be simplified as
\vspace{-0.15cm}
\begin{align}
\text{(N1.1):}~
P_{N1}\triangleq& \underset{{\bm \theta}}{\text{min}}
& &  \frac{(2^{\gamma_1}-1)2^{\gamma_2}{\sigma^2}}{\lambda_1({\bm \theta})}+\frac{(2^{\gamma_2}-1){\sigma^2}}{\lambda_2({\bm \theta})}\label{power_N1}\\
& \text{s.t.} & &  \theta_m \in {\cal F}, \forall m=1,\ldots,M\\
\text{(N2.1):}~
P_{N2}\triangleq& \underset{{\bm \theta}}{\text{min}}
& &  \frac{(2^{\gamma_1}-1){\sigma^2}}{\lambda_1({\bm \theta})}+\frac{(2^{\gamma_2}-1)2^{\gamma_1}{\sigma^2}}{\lambda_2({\bm \theta})}\label{power_N2}\\
& \text{s.t.} & & \theta_m \in {\cal F}, \forall m=1,\ldots,M.
\end{align}
Comparing \eqref{power_N1} and \eqref{power_N2}, we have
\vspace{-0.15cm}
\begin{align}
\hspace{-0.15cm}&\frac{(2^{\gamma_1}\hspace{-0.05cm}-\hspace{-0.05cm}1)2^{\gamma_2}{\sigma^2}}{\lambda_1({\bm \theta})}\hspace{-0.05cm}+\hspace{-0.05cm}\frac{(2^{\gamma_2}\hspace{-0.05cm}-\hspace{-0.05cm}1){\sigma^2}}{\lambda_2({\bm \theta})}\hspace{-0.05cm}-\hspace{-0.05cm}\frac{(2^{\gamma_1}\hspace{-0.05cm}-\hspace{-0.05cm}1){\sigma^2}}{\lambda_1({\bm \theta})}\hspace{-0.05cm}-\hspace{-0.05cm}\frac{(2^{\gamma_2}\hspace{-0.05cm}-\hspace{-0.05cm}1)2^{\gamma_1}{\sigma^2}}{\lambda_2({\bm \theta})}\notag\\
&=(2^{\gamma_1}-1)(2^{\gamma_2}-1){\sigma^2}\left( \frac{1}{\lambda_1({\bm \theta})}
-\frac{1}{\lambda_2({\bm \theta})}\right)
\end{align}
which is non-positive for $\lambda_1({\bm \theta}) \ge \lambda_2({\bm \theta})$ and 
non-negative for $\lambda_2({\bm \theta}) \ge \lambda_1({\bm \theta})$, respectively.
Thus, the minimum power required by using NOMA is obtained as 
\vspace{-0.2cm}
\begin{align}
\hspace{-0.2cm}P_{N}\triangleq\min\{P_{N1}, P_{N2}\}=\left\{ \begin{gathered}
P_{N1},\quad \lambda_1({\bm \theta}_N^*) \ge \lambda_2({\bm \theta}_N^*) \hfill \\
P_{N2},\quad {\rm otherwise} \hfill
\end{gathered}  \right.\hspace{-0.2cm}
\end{align}
\rev{with ${\bm \theta}_N^*$ being the optimal phase-shift vector.}

\vspace{-0.4cm}
\subsection{OMA Transmission Schemes}
\subsubsection{FDMA}
As shown in Fig.~\ref{system}, the AP communicates with two users \emph{simultaneously} over two equal frequency-domain adjacent RBs via FDMA.
Accordingly, the optimization problem of minimizing the total transmit power
at the AP is formulated as
\vspace{-0.2cm}
\begin{align}
\text{(F1):}~
P_{F}\triangleq& \underset{ {\bm \theta},P_1, P_2}{\text{min}}
& & P_1+P_2\\
& \text{~~s.t.} & & \frac{1}{2}\log_2\left(1+\frac{ P_1 \lambda_1({\bm \theta}) }{\frac{1}{2}\sigma^2} \right)\ge \gamma_1\label{con0_F1} \\
& & & \frac{1}{2}\log_2\left(1+\frac{ P_2  \lambda_2({\bm \theta}) }{\frac{1}{2}\sigma^2} \right) \ge \gamma_2\label{con1_F1}\\
& & & \theta_m \in {\cal F}, \forall m=1,\ldots,M.
\end{align}
Note that the factor $1/2$ in \eqref{con0_F1} and \eqref{con1_F1} is due to the fact that each user is assigned with half of the bandwidth as compared to the case of NOMA.  
Similar to NOMA, 
the inequality rate constraints can be replaced with equality constraints and problem (F1) is transformed into
\vspace{-0.2cm}
\begin{align}
\text{(F1.1):}~
P_{F}\triangleq& \underset{ {\bm \theta}}{\text{min}}
& & \frac{(2^{2\gamma_1}-1)\sigma^2}{2\lambda_1({\bm \theta})} + \frac{(2^{2\gamma_2}-1)\sigma^2}{2\lambda_2({\bm \theta})}\label{power_F1}\\
& \text{s.t.} & & \theta_m \in {\cal F}, \forall m=1,\ldots,M.
\end{align}

\subsubsection{TDMA} 
As shown in Fig.~\ref{system}, the AP communicates with two users \emph{consecutively} over two equal time-domain adjacent RBs via TDMA.
Different from NOMA and FDMA where \rev{the IRS phase-shift vector} needs to be set identical for the two users, in the case of TDMA \rev{the IRS phase-shift vector} can be set different for the two users over different time. This is fundamentally due to the hardware limitation of IRS passive reflection, which can be made \emph{time-selective}, but not \emph{frequency-selective} \cite{qingqing2019towards}.
Thus, the optimization problem of minimizing the total transmit power at the AP in the case of TDMA can be formulated as
\begin{align}
\hspace{-0.2cm}\text{(T1):}~
P_{T}\triangleq& \underset{ {\bm \theta}_1,{\bm \theta}_2,P_1, P_2}{\text{min}}
& & P_1+P_2 \\
& ~~~~\text{s.t.} & &  \hspace{-0.5cm}\frac{1}{2}\log_2\left(1+\frac{ 2P_1  \lambda_1({\bm \theta}_1) }{\sigma^2} \right)\ge \gamma_1\label{con0_T1} \\
& & & \hspace{-0.5cm}\frac{1}{2}\log_2\left(1+\frac{2P_2  \lambda_2({\bm \theta}_2) }{\sigma^2} \right) \ge \gamma_2\label{con1_T1}\\
& & & \hspace{-0.5cm}\theta_{k,m} \hspace{-0.1cm}\in\hspace{-0.1cm} {\cal F}, \forall m=1,\ldots,M, k\hspace{-0.1cm}\in\hspace{-0.1cm} \{1,2\}\hspace{-0.1cm}
\end{align}
where ${\bm \theta}_k \triangleq [{\theta_{k,1}},  {\theta_{k,2}},\ldots,{\theta_{k,M}}]^T$ denotes \rev{the phase-shift vector} for the $k$-th user.
Note that the factor $1/2$ in \eqref{con0_T1} and \eqref{con1_T1} is due to the fact that each user is assigned with half of the time as compared to the case of NOMA; as a result, there is an equivalent power gain of $2$ for each user shown in \eqref{con0_T1} and \eqref{con1_T1} for fair comparison with NOMA.   
Similar to (F1), (T1) can be transformed into
 \begin{align}
\hspace{-0.2cm} \text{(T1.1):}~
 P_{T}\triangleq& \underset{ {\bm \theta}_1}{\text{min}} ~~
   \frac{(2^{2\gamma_1}-1)\sigma^2}{2\lambda_1({\bm \theta}_1)} + \underset{ {\bm \theta}_2}{\text{min}} ~~\frac{(2^{2\gamma_2}-1)\sigma^2}{2\lambda_2({\bm \theta}_2)}\label{power_T1}\\
 & \text{s.t.}   ~~\theta_{k,m} \hspace{-0.05cm}\in\hspace{-0.05cm} {\cal F}, \forall m=1,\ldots,M, k\in \{1,2\}.\hspace{-0.1cm}
 \end{align}
 \vspace{-0.4cm}
\section{Performance Comparison and Low-Complexity Solution}
 \vspace{-0.2cm}
\subsection{Comparison of Minimum Transmit Power}\label{Comparison}
First, we compare the minimum transmit powers required by FDMA and TDMA for the IRS-assisted system, whose relationship is given by the following proposition.

\indent\emph{Proposition 1}: As the minimum of the sum of two functions is no less than the sum of their individual minimum values,
it follows that $P_{F}\ge P_{T}$ by comparing \eqref{power_F1} and \eqref{power_T1}, and the equality holds if and only if 
\begin{align}
\arg \underset{ {\bm \theta}\in {\cal F}^M }{\text{max}} ~~
\lambda_1({\bm \theta}) =\arg \underset{ {\bm \theta}\in {\cal F}^M }{\text{max}} ~~\lambda_2({\bm \theta}).
\end{align}
Note that the above result is a direct consequence of passive IRS reflection that can be \emph{time-selective}, but cannot be \emph{frequency-selective}. 

Next, we compare the minimum transmit powers required by FDMA and NOMA for the IRS-assisted system, with the result given by the following proposition.

\indent\emph{Proposition 2}: 
The minimum transmit powers required by
NOMA and FDMA satisfy $P_{F}\ge P_{N}$,
and the equality holds if and only if ${\bm \theta}_N^*={\bm \theta}_F^*$,
$\lambda_1({\bm \theta}_F^*)=\lambda_2({\bm \theta}_F^*)$, and $\gamma_1=\gamma_2$.
\begin{IEEEproof}
	Let ${\bm \theta}_F^*$ denote \rev{the optimal phase-shift vector} for (F1.1) associated with FDMA.
	For the case of $\lambda_1({\bm \theta}_F^*) \ge \lambda_2({\bm \theta}_F^*)$, the power gap between 
	\eqref{power_N1} and \eqref{power_F1} is given by
	\begin{align}
	&\Delta P_1= ~P_{F}-P_{N1}\notag\\
	\stackrel{(a)}{\ge}& 
	\frac{(2^{2\gamma_1}\hspace{-0.1cm}-\hspace{-0.1cm}1)\sigma^2}{2\lambda_1({\bm \theta}_F^*)} + \frac{(2^{2\gamma_2}\hspace{-0.1cm}-\hspace{-0.1cm}1)\sigma^2}{2\lambda_2({\bm \theta}_F^*)} -\frac{(2^{\gamma_1}\hspace{-0.1cm}-\hspace{-0.1cm}1)2^{\gamma_2}{\sigma^2}}{\lambda_1({\bm \theta}_F^*)}-\frac{(2^{\gamma_2}\hspace{-0.1cm}-\hspace{-0.1cm}1){\sigma^2}}{\lambda_2({\bm \theta}_F^*)}\notag\\
	= & \frac{( 
		2^{2\gamma_1}-1-2^{\gamma_1+\gamma_2+1 }+2^{\gamma_2+1}){\sigma^2}}{2\lambda_1({\bm \theta}_F^*)}+\frac{(2^{2\gamma_2}-2^{\gamma_2+1}+1){\sigma^2}}{2\lambda_2({\bm \theta}_F^*)}\notag \\
	\stackrel{(b)}{\ge}& 
	\frac{(2^{2\gamma_2}-
		2^{\gamma_1+\gamma_2+1 }+2^{2\gamma_1} 
		){\sigma^2}}{2\lambda_1({\bm \theta}_F^*)}\notag
	\\
	=& 
	\frac{(2^{\gamma_2}-2^{\gamma_1} 
		)^2{\sigma^2}}{2\lambda_1({\bm \theta}_F^*)}
	\stackrel{(c)}{\ge} 0 \label{difference1}
	\end{align}
	where the equality of $(a)$ holds if ${\bm \theta}_F^*$ is also an optimal solution to problem (N1.1); the equality of $(b)$ holds when 
	$\lambda_1({\bm \theta}_F^*)=\lambda_2({\bm \theta}_F^*)$; and the equality of $(c)$ holds when $\gamma_1=\gamma_2$.
%
Similarly, for the case of $\lambda_2({\bm \theta}_F^*) \ge \lambda_1({\bm \theta}_F^*)$, we can also obtain a non-negative power gap between \eqref{power_N2} and \eqref{power_F1}, i.e., $\Delta P_2= P_{F}-P_{N2}\ge 0$.
	Since $P_{F}\ge P_{N1}$ and $P_{F}\ge P_{N2}$, we can conclude that $P_{F}\ge \min\{P_{N1}, P_{N2}\}= P_{N}$, where the equality holds when the above conditions are all satisfied.
\end{IEEEproof}
\indent\emph{Remark 1}: From the above two propositions, we obtain $P_{F}\ge P_{T}$ and $P_{F}\ge P_{N}$, which implies that the minimum transmit power required by FDMA is always no lower than that of TDMA or NOMA.
However, there is no deterministic relationship between the minimum transmit power with NOMA versus that with TDMA, which generally depends on the 
locations and target rates of the two users, as will be shown later in Section \ref{Sim} by numerical examples. 
\vspace{-0.2cm}
\subsection{Optimal Solution}
Due to the discrete unit-modulus constraint on ${\bm \theta}$ and the non-convex objective functions, there is no standard method for efficiently solving the non-convex optimization problems (N1.1), (N2.1), (F1.1), and (T1.1) with globally optimal solutions. 
One straightforward approach is to search for all possible combinations of discrete
phase shifts
at all sub-surfaces and choose the one that achieves the
minimum transmit power. However, such a brute-force search method incurs an exponential complexity of ${\cal O}(L^M)$, which is prohibitively high for practical systems with large $M$ and/or $L$.
Although other approaches such as the branch-and-bound method can be applied to reduce the complexity \cite{wu2019beamforming},
the worst-case complexity is still exponential over $M$ due to the fundamental NP-hardness.
Thus, it mainly serves as a benchmark for evaluating other suboptimal schemes. 

\vspace{-0.2cm}
\subsection{Suboptimal Solution} \label{Suboptimal}
Note that the objective functions of \eqref{power_N1}, \eqref{power_N2} for NOMA, and \eqref{power_F1} for FDMA are
the weighed sum of two inverse channel power gains, which can be expressed in a generic form as
\vspace{-0.1cm}
\begin{align}\label{generic}
Q({\bm \theta})=\frac{a_1}{\lambda_1({\bm \theta})} + \frac{a_2}{\lambda_2({\bm \theta})}, \quad \theta_m \in {\cal F}, \forall m=1,\ldots,M
\end{align}
where $a_1$ and $a_2$ are two positive constants.
Dropping the discrete constraint, \rev{the phase-shift vector} that maximizes only each of the two user channel power gains is given by 
\vspace{-0.1cm}
\begin{align}
{\bm u}_k =\arg \underset{ {\bm \psi}}{\text{max}} ~~
\lambda_k({\bm \psi})=e^{j\angle h_{d,k}} e^{j\angle {\bm q}_k},\quad k\in \{1,2\}
\end{align}
where $\angle(\cdot)$ returns the phase of each element and
${\bm \psi}\triangleq[{\psi_1},  {\psi_2},\ldots,{\psi_M}]^T$ with $|{\psi_m}|=1 , \forall m=1,\ldots,M$.
Note that ${\bm u}_1\neq{\bm u}_2$ in general,
\rev{which implies that one phase-shift vector ${\bm \psi}$} cannot maximize two channel power gains $\lambda_1({\bm \psi})$ and $\lambda_2({\bm \psi})$ at the same time. 
As such, ${\bm \psi}$ needs to be properly designed to balance the channel power gains of two users for minimizing $Q({\bm \psi})$ in \eqref{generic}. 

To this end, we first define the non-negative linear combination of ${\bm u}_1$ and ${\bm u}_2$ as
\vspace{-0.1cm}
\begin{align}\label{segment}
\bar{\bm u}^{[\eta]}\triangleq \eta {\bm u}_1+ (1-\eta){\bm u}_2, \quad 0\le \eta \le 1.
\end{align}
As $\bar{\bm u}^{[\eta]}$ may not satisfy the unit-modulus constraint, we try to find its nearest 
unit-modulus vector $\tilde{\bm u}^{[\eta]}$, which is given by
\begin{align}
\tilde{\bm u}^{[\eta]} =\arg \underset{ {\bm \psi}}{\text{min}} ~~  \left\| {\bm \psi}-\bar{\bm u}^{[\eta]}\right\|^2=e^{j\angle \bar{\bm u}^{[\eta]}}, \quad 0\le \eta \le 1
\end{align}
to achieve a good balance between the two channel power gains by varying $\eta$. 
For simplicity, we divide $ [0, 1]$ into $B+1$ equal levels, which are denoted by the set $ {\cal {B}}=\left\{0, \frac{1}{B},\ldots, \frac{B-1}{B}, 1\right\}$.
Then, for each $\eta \in {\cal {B}}$, we quantize each phase shift in $\tilde{\bm u}^{[\eta]}$ to its nearest point in ${\cal F}$ to obtain \rev{the discrete phase-shift vector ${\bm \theta}^{[\eta]}$,} with each element given by
\begin{align}\label{quantized}
{\theta}^{[\eta]}_m=\arg \underset{ {\theta}\in {\cal F}}{\text{min}} ~~
\left|{\theta}-\tilde{ u}^{[\eta]}_m \right|^2, \quad m=1,\ldots,M.
\end{align}
Finally, we obtain the suboptimal solution ${\bm \theta}^{[\eta^*]}$ according to
\begin{align}\label{suboptimal}
\eta^*=\arg \underset{ \eta \in {\cal {B}} }{\text{min}} ~~ \frac{a_1}{\lambda_1({\bm \theta}^{[\eta]})} + \frac{a_2}{\lambda_2({\bm \theta}^{[\eta]})}.
\end{align}
From the above, we see that (N1.1), (N2.1), and (F1.1) can be solved suboptimally based on \eqref{quantized} and \eqref{suboptimal} with
a linear complexity of ${\cal O}(BML)$ only, which is thus referred to as the linear approximation (LA) method.
Furthermore, to further improve the solution obtained in \eqref{quantized} and \eqref{suboptimal}, the alternating optimization (AO) method \cite{wu2019beamforming} can be applied to solve \eqref{generic} based on the initial solution obtained by the LA method. Specifically, we alternately optimize one phase shift ${\theta}_m$ via one-dimensional
search over ${\cal F}$ while fixing the other $M-1$ phase shifts $\{{\theta}_i\}_{i \neq m}$, and iterate the above until the convergence is reached.
Note that the proposed algorithm is guaranteed to converge since
the objective value of \eqref{generic} is non-increasing over the iterations and lower-bounded by a
finite value $\frac{a_1}{\lambda_1({\bm u}_1)} + \frac{a_2}{\lambda_2({\bm u}_2)}$.
Given the number of iterations $I$, the total complexity
of the proposed algorithm is ${\cal O}((B+I)ML)$.

On the other hand, for (T1.1) associated with TDMA, since ${\bm \theta}_1$ and ${\bm \theta}_2$ are decoupled in the objective function of \eqref{power_T1}, we can simply quantize each phase shift in ${\bm u}_1$ and ${\bm u}_2$ to its nearest point in ${\cal F}$ to obtain \rev{discrete phase-shift vectors ${\bm \theta}_1^*$ and ${\bm \theta}_2^*$}, with each element given by
\begin{align}\label{sub-TDMA}
\hspace{-0.1cm} \theta_{k,m}^*=\arg \underset{ {\theta}\in {\cal F}}{\text{min}} ~
 \left|{\theta}\hspace{-0.05cm}-\hspace{-0.05cm} {u}_{k,m}\right|^2, m\hspace{-0.1cm}=\hspace{-0.1cm}1,\ldots,M, k\in \{1,2\}.\hspace{-0.2cm}
\end{align}
With the above for initialization, the AO method can be similarly applied to refine the solution to (T1.1). The overall complexity is thus of ${\cal O}((2+I)ML)$.

\vspace{-0.15cm}
\section{Simulation Results}\label{Sim}

\begin{figure}[!t]
	\centering
	\includegraphics[width=2.3in]{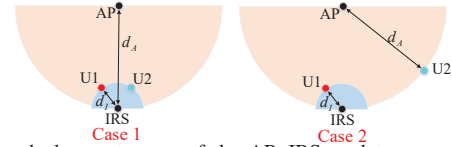}
	\caption{Two deployment cases of the AP, IRS, and two users (top view), where $d_A=50$ m and $d_I=4$ m.}
	\vspace{-0.6cm}
	\label{User_deployment}
\end{figure}

In this section, we present simulation results to numerically validate our analytical results (Propositions 1 and 2) and compare the performance of OMA (TDMA and FDMA) with NOMA in an IRS-assisted two-user system. The total number of reflecting elements of the IRS is set to $N=100$, which is divided into $M=5$ sub-surfaces.\footnote{The number of sub-surfaces is kept to be small  for implementing the brute-force search to obtain the optimal solution and provide the performance upper bound for the proposed algorithm.}
\rev{The path loss exponents of the AP$\rightarrow$users, IRS$\rightarrow$users, and AP$\rightarrow$IRS links are set as $3.2$, $2.6$, and $2.5$, respectively, and the path loss at the reference distance of 1 meter (m) is set as $30$ dB for each individual link \cite{Wu2019TWC,zheng2019intelligent,wu2019beamforming}.}
Moreover, the small-scale fading is characterized by Rayleigh fading for each individual link.
We consider two deployment cases shown in Fig.~\ref{User_deployment}, where $d_I=4$ m and $d_A=50$ m.
The results are averaged over $100$ independent fading channel realizations, with $L=8$ and ${\sigma^2}=-80$ dBm.

\begin{figure}
	\centering
	\hspace{-0.2cm}\subfigure[AP transmit power versus the common target rate $\gamma_0$.]{
		\begin{minipage}[b]{0.23\textwidth}
			\includegraphics[width=1.05\textwidth]{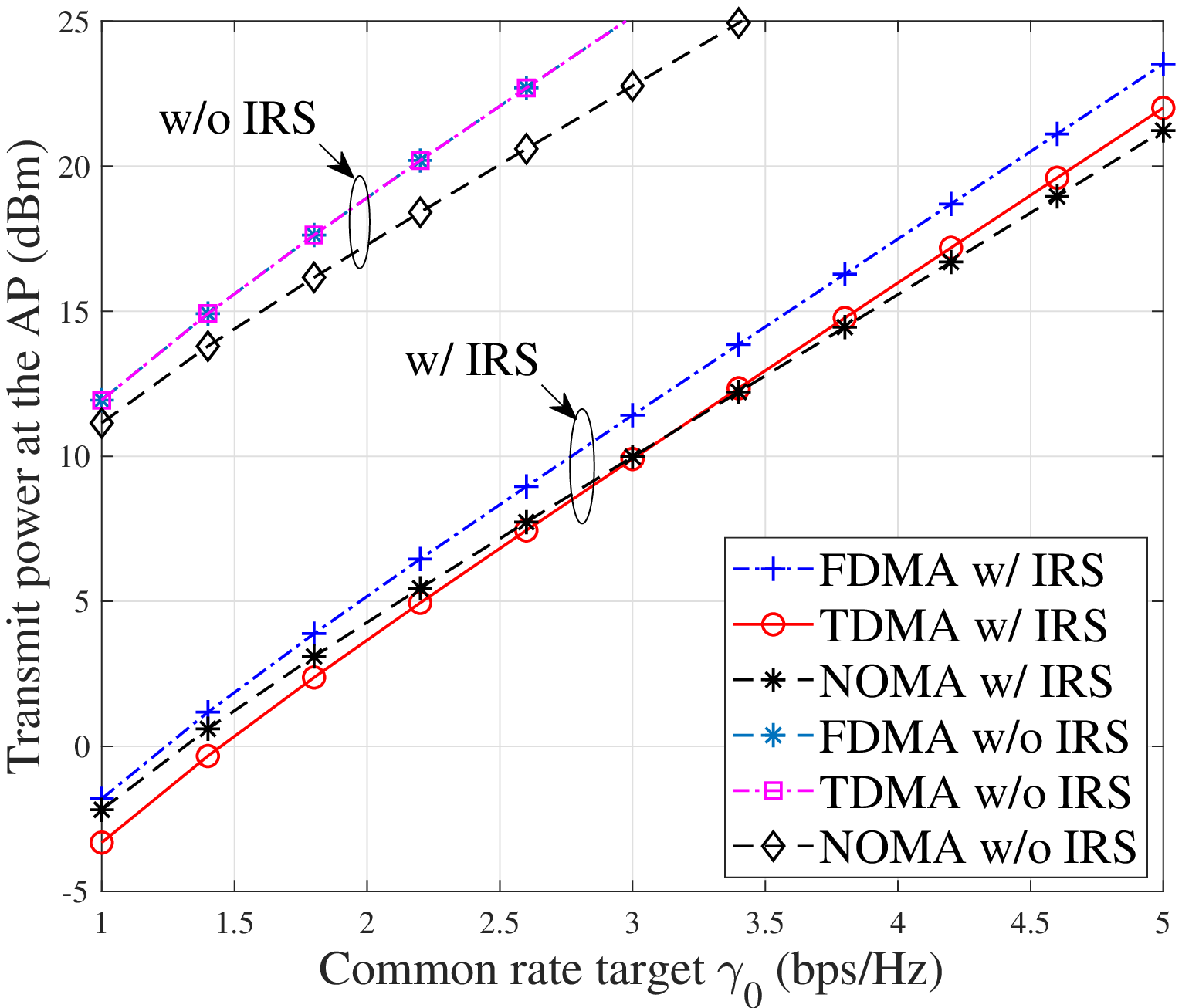}
		\end{minipage}\label{Result_type1}
	}
	\hspace{0.2cm}\subfigure[AP transmit power versus user 1's target rate $\gamma_1$, with the two users' sum rate fixed as $\gamma_1+\gamma_2=4$ bps/Hz.]{
		\begin{minipage}[b]{0.23\textwidth}
			\includegraphics[width=1.05\textwidth]{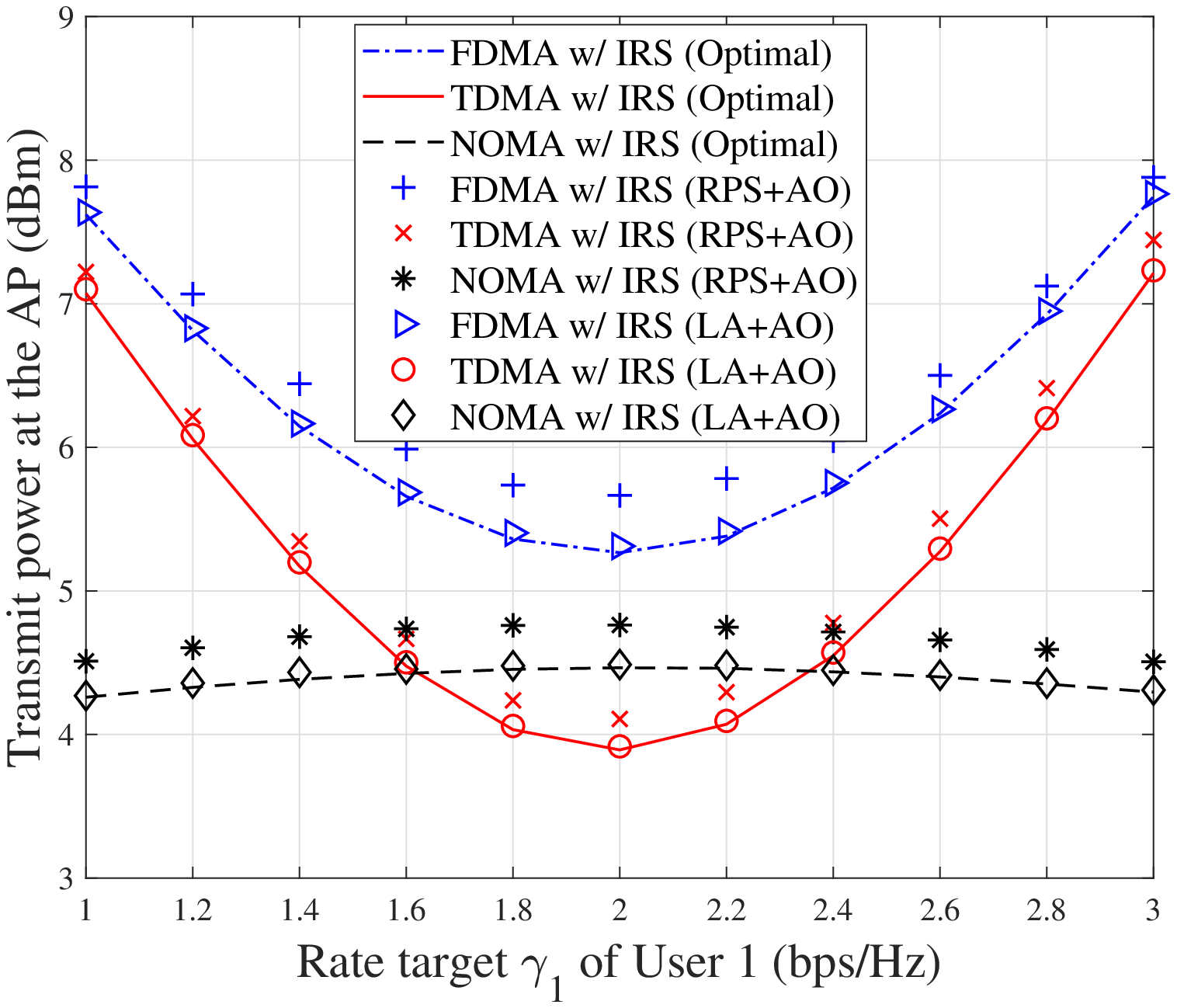}
		\end{minipage}\label{beta_type1}
	}
\setlength{\abovecaptionskip}{-8pt}
	\caption{Performance comparison of OMA and NOMA in Case 1.} \label{case1}
	\vspace{-0.6cm}
\end{figure}

We first consider Case 1 in Fig.~\ref{User_deployment},
where the two users are both located in the vicinity of the IRS with equal distances from the IRS as well as the AP.
In Fig.~\ref{Result_type1}, we compare the AP transmit powers required by different multiple access schemes versus the common target rate $\gamma_1=\gamma_2=\gamma_0$ for the two users.
Apparently, compared to the schemes without IRS, the required
transmit power at the AP is significantly reduced with the aid of IRS.
Moreover, one can observe that without IRS, the required
transmit power by NOMA is always lower than that of OMA (same for TDMA and FDMA).
\rev{In contrast, when the IRS is deployed to assist two users in its vicinity, the above conclusion is not valid in general.
For example, we observe that the required transmit power by TDMA is lower than that of NOMA for the common target rate $\gamma_0$ less than 3~bps/Hz. This is due to the use of different IRS phase-shift designs for the two users by exploiting the IRS's time selectivity via TDMA.}
Moreover, FDMA always requires higher transmit power than the other two counterparts, which is in accordance with Propositions 1 and 2.
On the other hand, by increasing the common target rate $\gamma_0$, the power growth rate with TDMA is slightly higher than that with NOMA,
which can be well understood since the required power by TDMA in \eqref{power_T1} is the sum of two exponential functions with a higher exponent of $2\gamma_0$, as compared to that in \eqref{power_N1} or \eqref{power_N2} with NOMA.


To draw more insight for Case~1, we depict the AP transmit powers required by different transmission schemes versus user~1's target rate $\gamma_1$ in Fig.~\ref{beta_type1}, with the two users' sum rate fixed as $\gamma_1+\gamma_2=4$ bps/Hz.
First, we can observe that TDMA requires lower transmit power than NOMA when the target rates of the two users are close (i.e., symmetric).
Second, the transmit powers of both TDMA and FDMA are observed to dramatically increase when the two users' rates become more \emph{asymmetric}, which is attributed to the exponentially increasing of the dominant user target rate, i.e., $\text{max}\{\gamma_1,\gamma_2\}$
in \eqref{power_F1} and \eqref{power_T1}.
In contrast, it is observed that the required transmit power by NOMA is almost insensitive to the disparity of user target rates, thus providing a more robust performance.
Third, to show the performance of the proposed suboptimal solution in Section~\ref{Suboptimal} (i.e., the LA method with $B=8$ followed by the AO method with $I=2$), we consider two benchmark schemes: 1) the optimal solution by the brute-force search and 2) the random phase shift (RPS) initialization followed by the AO method with $10$ iterations (which has the same complexity as the proposed one for fair comparison).
One can observe that our proposed solution achieves near-optimal performance, whereas 
a non-negligible performance loss is observed for the RPS-based initialization as compared to the optimal solution. The above result shows the effectiveness of the low-complexity LA method.  

\begin{figure}
	\centering
	\hspace{-0.2cm}\subfigure[AP transmit power versus the common target rate $\gamma_0$.]{
		\begin{minipage}[b]{0.23\textwidth}
			\includegraphics[width=1.05\textwidth]{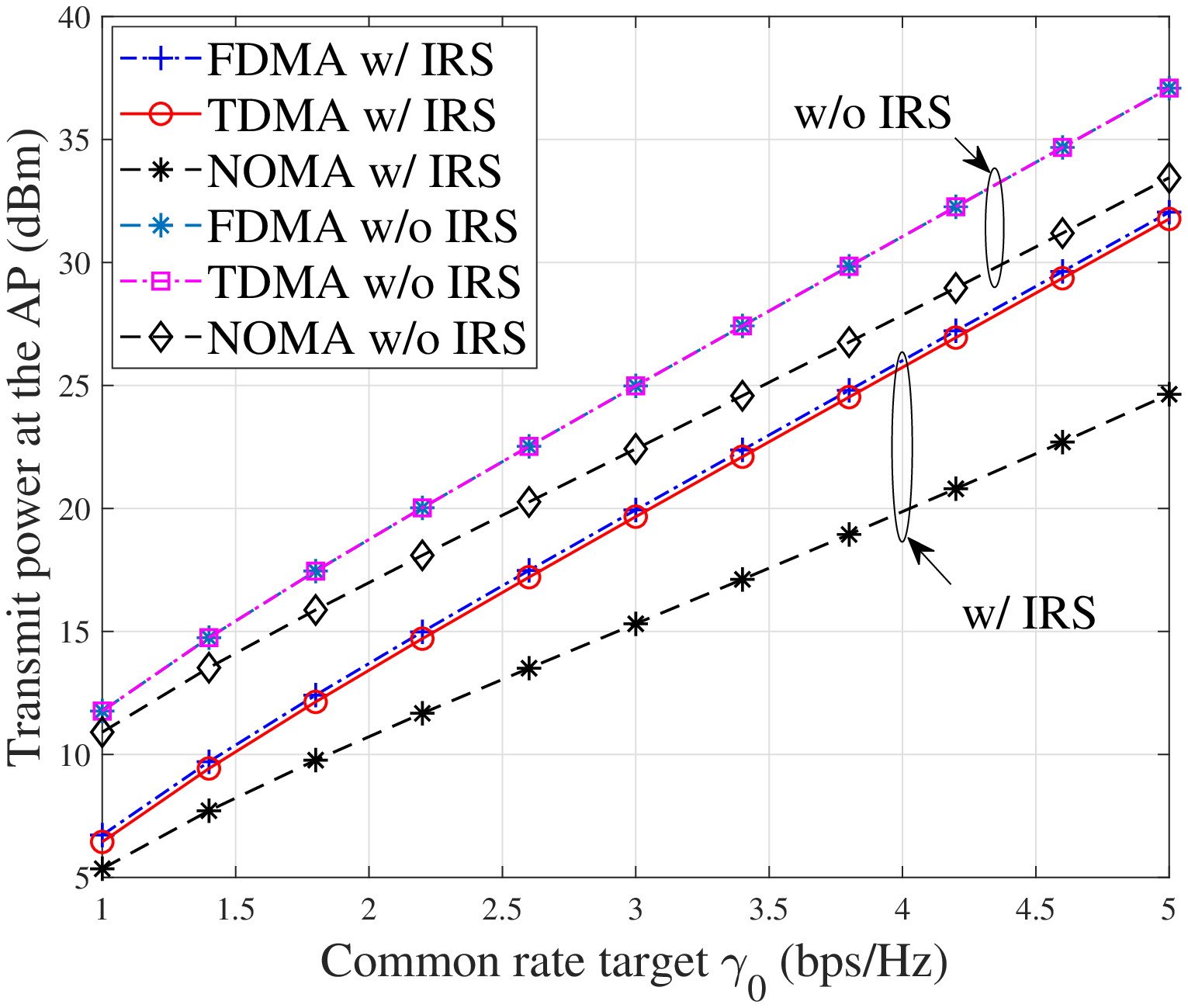}
		\end{minipage}\label{Result_type2}
	}
	\hspace{0.2cm}\subfigure[AP transmit power versus user~1's target rate $\gamma_1$, with the two users' sum rate fixed as $\gamma_1+\gamma_2=4$ bps/Hz.]{
		\begin{minipage}[b]{0.23\textwidth}
			\includegraphics[width=1.05\textwidth]{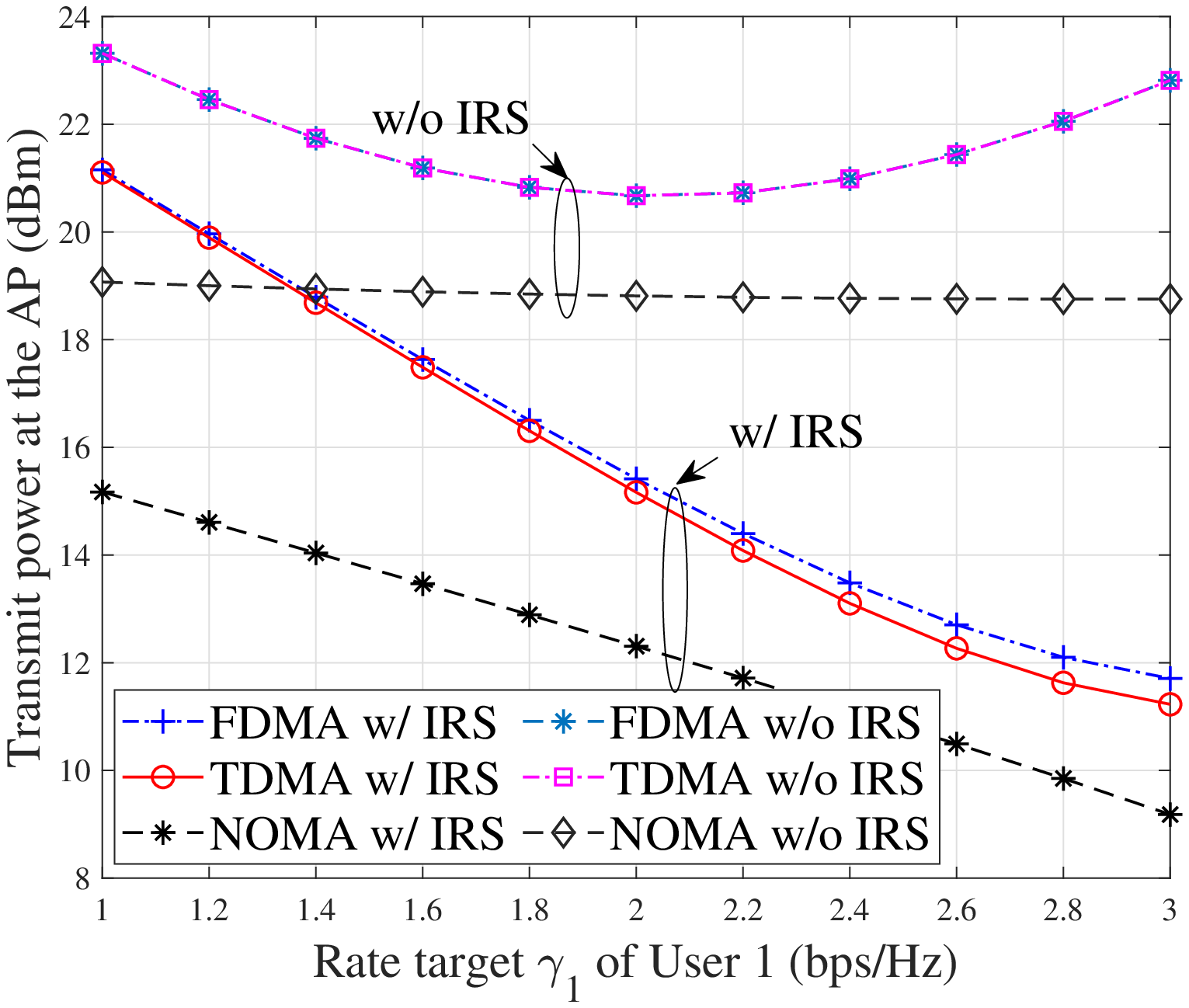}
		\end{minipage}\label{beta_type2}
	}
\setlength{\abovecaptionskip}{-8pt}
	\caption{Performance comparison of OMA and NOMA in Case 2.}\hspace{-0.8cm} \label{case2}
	\vspace{-0.8cm}
\end{figure}

Next, unlike Case 1 with two near-IRS users (namely, \emph{symmetric} user deployment), we consider another setup of Case 2 with one near-IRS user and one far-IRS user (namely, \emph{asymmetric} user deployment).
As one user moves far away from the IRS and thus needs more power directly from the AP, it is observed in Figs.~\ref{Result_type2} and \ref{beta_type2} that the transmit power required at the AP increases drastically compared to Figs.~\ref{Result_type1} and~\ref{beta_type1}.
This is expected since it is more energy efficient when both users are located
in the vicinity of the IRS to reap its passive beamforming gain for saving the AP transmit power.
Moreover, one can observe from Figs.~\ref{Result_type2} and \ref{beta_type2} that the required
transmit power by NOMA is always lower than that by OMA, regardless of whether TDMA or FDMA is used.
This can be explained by the fact that NOMA has higher spectrum efficiency than OMA under \emph{asymmetric} user channels, even with IRS deployed to assist one of the two users.
Furthermore, since the IRS reflection has negligible effect on the far-IRS user,
TDMA and FDMA show nearly the same performance in the IRS-assisted system.


\vspace{-0.25cm}
\section{Conclusions}
In this letter, we have optimized the IRS reflection with \emph{discrete phase shifts} for the IRS-assisted NOMA and OMA systems to minimize the AP transmit power with given user rates.
The minimum transmit powers required by different multiple access schemes have been compared analytically and a low-complexity algorithm has been proposed to achieve near-optimal performance.
\rev{In particular, our work has revealed that NOMA may perform worse than TDMA for near-IRS users with \emph{symmetric} rates.
This thus provides an important guideline for user paring in IRS-aided systems with a large number of users and RBs (e.g., IRS-assisted orthogonal FDMA (OFDMA)): it is preferable to pair users with \emph{asymmetric} rates and/or \emph{asymmetric} deployment (i.e., with distinct distances from the IRS, regardless of the distances from the AP) to exploit the NOMA gain over OMA.}


%
%
%
%
%
%
%
%

\ifCLASSOPTIONcaptionsoff
  \newpage
\fi
\vspace{-0.25cm}
\bibliographystyle{IEEEtran}
\bibliography{IRS_NOMA}

\begin{thebibliography}{10}
\providecommand{\url}[1]{#1}
\csname url@samestyle\endcsname
\providecommand{\newblock}{\relax}
\providecommand{\bibinfo}[2]{#2}
\providecommand{\BIBentrySTDinterwordspacing}{\spaceskip=0pt\relax}
\providecommand{\BIBentryALTinterwordstretchfactor}{4}
\providecommand{\BIBentryALTinterwordspacing}{\spaceskip=\fontdimen2\font plus
\BIBentryALTinterwordstretchfactor\fontdimen3\font minus
  \fontdimen4\font\relax}
\providecommand{\BIBforeignlanguage}[2]{{%
\expandafter\ifx\csname l@#1\endcsname\relax
\typeout{** WARNING: IEEEtran.bst: No hyphenation pattern has been}%
\typeout{** loaded for the language `#1'. Using the pattern for}%
\typeout{** the default language instead.}%
\else
\language=\csname l@#1\endcsname
\fi
#2}}
\providecommand{\BIBdecl}{\relax}
\BIBdecl

\bibitem{qingqing2019towards}
Q.~Wu and R.~Zhang, ``Towards smart and reconfigurable environment: Intelligent
  reflecting surface aided wireless network,'' \emph{IEEE Commun. Mag.}, doi:
  10.1109/MCOM.001.1900107, Nov. 2019.

\bibitem{Renzo2019Smart}
M.~Di~Renzo \emph{et~al.}, ``Smart radio environments empowered by
  reconfigurable {AI} meta-surfaces: An idea whose time has come,''
  \emph{EURASIP J. Wireless Commun. Netw.}, vol. 2019:129, May 2019.

\bibitem{Wu2019TWC}
Q.~Wu and R.~Zhang, ``Intelligent reflecting surface enhanced wireless network
  via joint active and passive beamforming,'' \emph{IEEE Trans. Wireless
  Commun.}, vol.~18, no.~11, pp. 5394--5409, Nov. 2019.

\bibitem{Gang2019intelligent}
G.~Yang, X.~Xu, and Y.-C. Liang, ``Intelligent reflecting surface assisted
  non-orthogonal multiple access,'' \emph{arXiv preprint arXiv:1907.03133},
  2019.

\bibitem{ding2019simple}
Z.~Ding and H.~Poor, ``A simple design of {IRS-NOMA} transmission,''
  \emph{arXiv preprint arXiv:1907.09918}, 2019.

\bibitem{li2019joint}
Y.~Li, M.~Jiang, Q.~Zhang, and J.~Qin, ``Joint beamforming design in
  multi-cluster {MISO NOMA} intelligent reflecting surface-aided downlink
  communication networks,'' \emph{arXiv preprint arXiv:1909.06972}, 2019.

\bibitem{mu2019exploiting}
X.~Mu, Y.~Liu, L.~Guo, J.~Lin, and N.~Al-Dhahir, ``Exploiting intelligent
  reflecting surfaces in multi-antenna aided {NOMA} systems,'' \emph{arXiv
  preprint arXiv:1910.13636}, 2019.

\bibitem{zheng2019intelligent}
B.~Zheng and R.~Zhang, ``Intelligent reflecting surface-enhanced {OFDM}:
  Channel estimation and reflection optimization,'' \emph{IEEE Wireless Commun.
  Lett.}, doi: 10.1109/LWC.2019.2961357, Dec. 2019.

\bibitem{yang2019intelligent}
Y.~Yang, B.~Zheng, S.~Zhang, and R.~Zhang, ``Intelligent reflecting surface
  meets {OFDM}: Protocol design and rate maximization,'' \emph{arXiv preprint
  arXiv:1906.09956}, 2019.

\bibitem{wu2019beamforming}
Q.~Wu and R.~Zhang, ``Beamforming optimization for wireless network aided by
  intelligent reflecting surface with discrete phase shifts,'' \emph{IEEE
  Trans. Commun.}, doi: 10.1109/TCOMM.2019.2958916, Dec. 2019.

\end{thebibliography}

\end{document}